\documentclass[aps,pra,twocolumn,superscriptaddress,showpacs]{revtex4-1}
\usepackage{bm}
\usepackage{amsthm}
\usepackage{amsmath,amssymb}
\usepackage{graphicx}
\usepackage{braket}
\usepackage{enumerate}
\usepackage{ascmac}
\usepackage{mathrsfs}

\newcommand{\e}[1]{\begin{align} #1\end{align}}

\newcommand{\tr}[1]{\mathrm{tr}\left[ #1 \right]}
\newcommand{\h}[1]{\hat{#1}}
\newcommand{\pd}[3]{
 \if 1#1 \frac{\partial #2}{\partial #3}
 \else \frac{\partial^{#1} #2}{\partial #3^{#1}}\fi}
 \newcommand{\od}[3]{
 \if 1#1 \frac{{\mathrm d} #2}{{\mathrm d} #3}
 \else \frac{{\mathrm d}^{#1} #2}{{\mathrm d}#3^{#1}}\fi}

\newcommand{\ginprod}[3]{\langle\hspace{-0.2em}\langle #1 \rangle\hspace{-0.2em}\rangle_{#2}^{#3}}
\newcommand{\ex}[0]{{\mathrm e}}

\begin{document}

%\preprint{}

\title{Determining the Continuous Family of Quantum Fisher Information from Linear Response Theory}


\author{Tomohiro Shitara}
\email[]{shitara@cat.phys.s.u-tokyo.ac.jp}
\affiliation{Department of Physics, University of Tokyo, Hongo 7-3-1, Bunkyo-ku, Tokyo 113-8654, Japan}
\author{Masahito Ueda}
\affiliation{Department of Physics, University of Tokyo, Hongo 7-3-1, Bunkyo-ku, Tokyo 113-8654, Japan}
\affiliation{RIKEN Center for Emergent Matter Science (CEMS), Wako 351-0198, Japan}
%\thanks{}


\date{\today}

\begin{abstract}
The quantum Fisher information represents the continuous family of metrics on the space of quantum states and places the fundamental limit on the accuracy of quantum state estimation.
We show that the entire family of the quantum Fisher information can be determined from linear response theory through generalized covariances.
We derive the generalized fluctuation-dissipation theorem that relates the linear response function to generalized covariances and hence allows us to determine the quantum Fisher information from linear response functions, which is experimentally measurable quantities.
As an application, we examine the skew information, which is one of the quantum Fisher information, of a harmonic oscillator in thermal equilibrium, and show that the equality of the skew information-based uncertainty relation holds.

\end{abstract}

%\keywords{}

\maketitle


%------------------------------------------------------------------------------------%
%------------------------------------------------------------------------------------%
\section{introduction}
It is vital to identify the natural metric on the space of physical states and to unveil their operational meanings.
In classical theory, where states are represented by probability distributions on phase space, the natural metric is uniquely determined by the monotonicity under information processing~\cite{Cencov1982} to be the Fisher metric or the Fisher information.
Here, the monotonicity means that the Fisher information decreases monotonically under any stochastic mapping.
The Fisher information gives the fundamental upper bound on the accuracy of state estimation through the Cram\'er-Rao inequality~\cite{Cramer1946}.
The Fisher metric also appears in nonequilibrium thermodynamics~\cite{Crooks2007, Sivak2012, Nakagawa2014}.
For example, in the linear response regime, the excess work needed to control thermal states can be expressed in terms of the Fisher metric on the space of equilibrium states~\cite{Sivak2012}.

In quantum theory, the noncommutativity of density operators results in the breakdown of the uniqueness of the Fisher information, and various types of quantum counterparts of the Fisher information can be considered.
The symmetric logarithmic derivative (SLD) Fisher information, which is the most familiar to physicists, was first introduced in quantum estimation theory~\cite{Helstrom1967} to place an upper bound of the accuracy of state estimation via the quantum Cram\'er-Rao inequality, and is often used to assess the efficiency of the quantum metrology~\cite{Giovannetti2006,Giovannetti2011}.
The SLD Fisher information is equivalent to the fidelity susceptibility up to a constant factor, and has been applied to condensed matter physics to analyze quantum phase transitions in the ground state~\cite{You2007,CamposVenuti2007,Zanardi2007} and in dissipative systems~\cite{Banchi2014,Marzolino2014}.
Furthermore, multipartite entanglement can be detected by the SLD Fisher information~\cite{Hyllus2012,Toth2012}.
However, there are diverse situations in which other types of the quantum Fisher information are relevant.
The right logarithmic derivative (RLD) Fisher information, for example, gives a tighter bound than the SLD Fisher information when we estimate multiple parameters of Gaussian systems~\cite{Yuen1973}.
The skew information~\cite{Wigner1963}, which is closely related to the Wigner-Araki-Yanase theorem~\cite{Wigner1952, Araki1960, Yanase1961} and uncertainty relations~\cite{Luo2005,Gibilisco2007a,Yanagi2010,Furuichi2010,Yanagi2011}, is yet another example of the quantum Fisher information.
The Bogoliubov-Kubo-Mori (BKM) Fisher information becomes relevant when we consider the distinguishability of two different states from the  viewpoint of large deviation~\cite{Hiai1991, Ogawa2000, Hayashi2002}.
All the quantum Fisher information mentioned above satisfy the monotonicity under quantum operations.
Indeed, if we regard the quantum Fisher information as the monotone metrics on the space of quantum states, there is a one-to-one correspondence between the quantum Fisher information and operator monotone functions~\cite{Petz1996}.
However, neither a unified understanding of operational meanings of  various quantum Fisher information contents nor how to experimentally determine the general quantum Fisher information is known.

In this paper, we develop a method to determine all types of the quantum Fisher information by exploiting the similarity between estimation theory (statistics) and linear response theory (statistical mechanics).
Since the quantum Fisher information represents the sensitivity, or the response, to infinitesimal changes of parameters that characterize the quantum state, it is quantitatively related to the correlation function, or the covariance, of the estimated value via the quantum Cram\'er-Rao inequality.
When the state is in equilibrium, such a covariance is quantitatively related to the response to an external perturbation, the fact known as the fluctuation-dissipation theorem~\cite{Kubo1957}.
By using these two connections, we can express the quantum Fisher information in terms of response functions.

% figures should be put into the text as floats.
% Use the graphics or graphicx packages (distributed with LaTeX2e)
% and the \includegraphics macro defined in those packages.
% See the LaTeX Graphics Companion by Michel Goosens, Sebastian Rahtz,
% entire page. There is no need to do explicit centering.

\begin{figure}
\begin{center}
\includegraphics[width = 0.9\columnwidth]{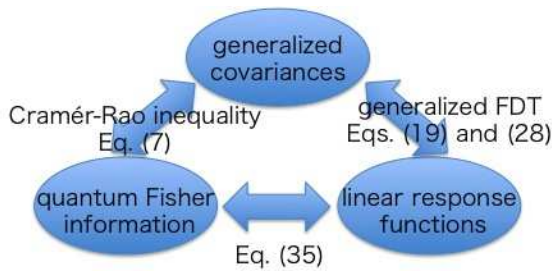}
\caption{
The relation between the generalized covariances, the quantum Fisher information, and linear response functions.
The quantum Fisher information gives lower bound of the generalized covariance of the estimator through the Cram\'er-Rao inequality~(7).
We derive the generalized fluctuation-dissipation theorem (FDT)~(19) and (28), which connects the generalized covariance and linear response functions.
Based on this theorem, we obtain the information-dissipation relation (35), which enables us to determine the quantum Fisher information from linear response functions, which is experimentally measurable.
\label{fig1}}
\end{center}
\end{figure}

What has been known about the relation between the linear response theory and the quantum Fisher information is that the static susceptibility is given by the canonical correlation, which is one of the generalized covariances corresponding to the BKM Fisher information~\cite{Petz1993}.
If we consider only the static susceptibility, however, we obtain the information on the canonical correlation and hence on the BKM Fisher information, and no information on other quantum Fisher information.
Therefore we need to consider dynamical response functions to relate the general quantum Fisher information to linear response functions.

This paper is organized as follows.
In Sec.~\ref{QFIandGC}, we introduce the quantum Fisher information and the generalized covariances.
In Sec.~\ref{LRT}, we briefly review the linear response theory.
In Sec.~\ref{GFDT}, we derive a generalized fluctuation-dissipation theorem, which relates generalized covariances to linear response functions such as the admittance and the dynamical susceptibility.
Based on this theorem, we show in Sec.~\ref{determine} how to determine the quantum Fisher information from the linear response to an external field.
In Sec.~\ref{SIUR}, we apply our method to the skew information-based uncertainty relations.
We discuss the applicability and the efficiency of our method in Sec.~\ref{discussion}.
Finally, we conclude the paper in Sec.~\ref{conclusion}.

%------------------------------------------------------------------------------------%
%------------------------------------------------------------------------------------%
\section{quantum Fisher information and generalized covariance}\label{QFIandGC}
Consider a family of density operators $\{\h\rho_{\bm\theta}\}$ parametrized by a real vector $\bm\theta\in\mathbb R^m$.
The state of the system is given by $\h\rho_{\bm\theta}$ for some fixed but unknown $\bm\theta$.
Our task is to perform an appropriate measurement to estimate $\bm\theta$ as accurately as possible.
When the estimation is unbiased, the variance of estimated parameters is larger than the inverse of the quantum Fisher information due to the quantum Cram\'er-Rao inequality.
Here, the quantum Fisher information is defined by
\e{
[J_{\bm\theta}^{\rm Q}]_{\mu\nu}=\tr{\pd{1}{\h\rho_{\bm\theta}}{\theta_\mu}(\bm K^f_{\h\rho_{\bm\theta}})^{-1}(\pd{1}{\h\rho_{\bm\theta}}{\theta_\nu})},
}
where $\bm K^f_{\h\rho}$ is a superoperator defined as $\bm K^f_{\h\rho}:=f(\bm L_{\h\rho}\bm R_{\h\rho}^{-1})\bm R_{\h\rho}$, and $f:(0,\infty)\rightarrow (0,\infty)$ is an operator monotone function satisfying $f(1)=1$;
$\bm R_{\h\rho} $ and $\bm L_{\h\rho}$ are superoperators that multiply $\h\rho$ from the right and the left, respectively:
\e{
\bm R_{\h\rho}(\h A):=\h A\h\rho, \ \bm L_{\h\rho}(\h A):=\h\rho\h A.
}
We note that the quantum Fisher information is defined for each operator monotone function $f(x)$, and characterized by the monotonicity under completely positive and trace-preserving mappings~\cite{Petz1996}.
The SLD Fisher information, which is the most familiar to physicists, corresponds to $f(x)=(x+1)/2$.
The quantum Cram\'er-Rao inequality immediately follows from the monotonicity and the classical Cram\'er-Rao inequality.

The key quantity that connects the quantum Fisher information with the response function is the generalized covariance~\cite{Petz2002, Gibilisco2009}.
The generalized covariance of two observables $\h X$ and $\h Y$ is defined as
\e{
\ginprod{\h X,\h Y}{\h\rho}{f}:=\tr{\h X^\dagger\bm K^f_{\h\rho}\h Y }.
}
It is a generalization of the classical covariance for two observables that do not necessarily commute with the density operator $\h\rho$.
We note that the operational meaning of the generalized covariance in the noncommutative case has not yet been fully understood, which corresponds to the fact that the operational meaning of the general quantum Fisher information has not been clarified yet.
The symmetrized correlation and the canonical correlation correspond to $f(x)=\frac{1+x}{2}$ and $f(x)=\frac{x-1}{\log x}$, respectively.
Then, using the logarithmic derivative $\h L_\mu:=(\bm K^f_{\h\rho_{\bm\theta}})^{-1}(\pd{1}{\h\rho_{\bm\theta}}{\theta_\mu})$, the quantum Fisher information can be rewritten as 
\e{
[J_{\bm\theta}^{\rm Q}]_{\mu\nu}=\ginprod{\h L_\mu,\h L_\nu}{\h\rho_{\bm\theta}}{f}. \label{QFIGC}
}
Defining the dual of an operator monotone function  $f(x)$ as $\tilde{f}(x):=xf(1/x)$, which is also operator monotone, we obtain
\e{
\ginprod{\h X,\h Y}{\h\rho}{f}=\ginprod{\h Y,\h X}{\h\rho}{\tilde{f}}.
}
In particular, if $f(x)=\tilde{f}(x)$ is satisfied, $f(x)$ is called standard, and the corresponding generalized covariance and the quantum Fisher information take real values.

%------------------------------------------------------------------------------------%
%------------------------------------------------------------------------------------%
The generalized covariances and the quantum Fisher information satisfy another type of the quantum Cram\'er-Rao inequality.
To see this, we consider a one-parameter model described as
\e{
\h\rho_\theta=\h\rho_0+\theta\pd{1}{\h\rho_{\theta}}{\theta}\Bigg|_{\theta=0}+O(\theta^2).
}
Suppose that we estimate the parameter $\theta$ by measuring an observable $\h O$.
If we impose a local unbiasedness at $\theta=0$, i.e., $\tr{\h\rho_\theta\h O}=\theta+O(\theta^2)$, we have
\e{
\ginprod{\h O,\h O}{\h\rho_0}{f}\ge \frac{1}{J^Q_{\theta=0}}
}
from the Schwarz inequality, where the equality is achieved for 
\e{
\h O=\frac{\h L}{\ginprod{\h L,\h L}{\h\rho_{\bm\theta}}{f}}.
}
In this sense, the quantum Fisher information gives the minimum value of the corresponding generalized covariance of operators with which we estimate $\theta$.

%------------------------------------------------------------------------------------%
%------------------------------------------------------------------------------------%
\section{Linear Response theory}\label{LRT}
Next we briefly review the linear response theory.
For a given state $\h\rho$, we assume that the system evolves under the Hamiltonian $\h H=-\frac{1}{\beta}\log\h\rho$, where $\beta=1/k_{\rm B}T$ is  the inverse temperature.
Or, equivalently, we consider the canonical ensemble for a given Hamiltonian $\h H$.
Then, we have
\e{
&\h H=\sum_iE_i\ket i\bra i,
\h\rho=\sum_i p_i \ket i\bra i,\\
&p_i=\ex^{-\beta(E_i-F)},
}
where $\ket i$'s are the eigenvectors of the density operator, and $F:=-\frac{1}{\beta}\log\tr{\ex^{-\beta\h H}}$ is the free energy.
In linear response theory,  if we apply an external perturbation $\h H^{\rm ext}(t)=-\sum_\nu X_\nu(t)\h A_\nu$, the expectation values of the displacement operator $\h A_\mu$ and the current operator $\h J_\mu(t):=\dot{\h A}_\mu(t)=\frac{1}{i\hbar}[\h A_\mu(t),\h H]$ deviate from the equilibirium as
\e{
\braket{\h J_\mu(t)}=\int_{-\infty}^t {\rm d}t' \sum_\nu \Phi_{\mu\nu}(t-t')X_\nu(t'),\\
\braket{\Delta\h A_\mu(t)}=\int_{-\infty}^t {\rm d}t' \sum_\nu \tilde{\Phi}_{\mu\nu}(t-t')X_\nu(t'),
}
where the coefficients $\Phi_{\mu\nu}(t-t')$ and $\tilde{\Phi}_{\mu\nu}(t-t')$ are called the linear response functions.
According to the Kubo formula~\cite{Kubo1957}, the linear response function is given by the commutator of the perturbation operator and the Heisenberg operator to be measured:
\e{
\Phi_{\mu\nu}(t)=\frac{1}{i\hbar}\tr{\h\rho[ \h A_\nu(0) , \h J_\mu(t) ]},\\
\tilde{\Phi}_{\mu\nu}(t)=\frac{1}{i\hbar}\tr{\h\rho[ \h A_\nu(0) , \h A_\mu(t) ]}.\label{Kubo2}
}
Furthermore, the linear response function of the current operator can also be written in terms of the canonical correlation of the current operators as 
\e{
\Phi_{\mu\nu}(t)=\beta \int_0^1 \tr{\h\rho^\lambda \h J_\mu(t) \h\rho^{1-\lambda} \h J_\nu(0)}{\rm d}\lambda.\label{Kubo}
}
In the frequency domain, the fluctuation-dissipation theorem holds, which states that the symmetrized correlation 
\e{
C_{\mu\nu,\omega}^{\rm sym}:=\frac{1}{2}\int^\infty_{-\infty}{\rm d}t\ex^{i\omega t}\tr{\h\rho(\h J_\mu(t)\h J_\nu(0)+\h J_\nu(0)\h J_\mu(t))}
}
is proportional to the Fourier transform of the response response function,
\e{
C_{\mu\nu,\omega}^{\rm sym}=\hbar\omega\cdot\frac{1}{2}\coth \left( \frac{\beta\hbar\omega}{2}\right)\Phi_{\mu\nu,\omega},  \label{FDT}
}
where the coefficient is equal to the energy of the harmonic oscillator in thermal equilibrium.

%@@@@@@@@@@@@temporal correlation function$B$OB,DjJ}K!$K6/$/0MB8$9$k$3$H$KCm0U(B[fujikura]$B!*(B
%------------------------------------------------------------------------------------%
%------------------------------------------------------------------------------------%
\section{Generalized fluctuation-dissipation theorem}\label{GFDT}
Now we are in a position to describe our main results of this paper.
We define the Fourier transform of the generalized covariance of two temporal current operators as
\e{
C_{\mu\nu,\omega}^f&:=\int_{-\infty}^\infty{\rm d}(t-t')\ex^{i(t-t')\omega}\ginprod{\h J_\mu(t), \h J_\nu(t')}{\h\rho}{f}.
}
Then, the fluctuation-dissipation theorem~\eqref{FDT} is generalized as
\e{
C_{\mu\nu,\omega}^f&=\hbar\omega\frac{f(\ex^{-\beta\hbar\omega})}{1-\ex^{-\beta\hbar\omega}}\Phi_{\mu\nu,\omega}. \label{gFDT}
}
In the frequency domain, any generalized covariance is proportional to the response $\Phi_{\mu\nu,\omega}$, and the choice of an operator monotone function $f(x)$ determines its frequency-dependence of the coefficient.
The usual fluctuation-dissipation theorem~\eqref{FDT} is reproduced by setting $f(x)=\frac{1+x}{2}$ in Eq.~\eqref{gFDT}, which corresponds to the SLD Fisher information.
We note that generalized covariances do not depend on $f(x)$ in the classical limit or the high-temperature limit $\beta\hbar\omega\rightarrow0$.
The term $\frac{f(\ex^{-\beta\hbar\omega})}{1-\ex^{-\beta\hbar\omega}}$ in Eq.~\eqref{gFDT} can be rewritten in terms of the expectation value of the number operator of the harmonic oscillator, $\bar{n}:=\frac{1}{\ex^\alpha-1}$ with $\alpha:=\beta\hbar\omega$.
Actually, it is equal to a generalized mean~\cite{Kubo1980} of $\bar{n}$ and $\bar{n}+1$ defined as $(\bar{n}+1)f(\frac{\bar{n}}{\bar{n}+1})$.
For example, $f(x)=\frac{x+1}{2}, \sqrt x, \frac{2x}{x+1}$ and $\frac{x-1}{\log x}$ correspond to the arithmetic, geometric, harmonic and logarithmic means, respectively.
The coefficients $\frac{f(\ex^{-\beta\hbar\omega})}{1-\ex^{-\beta\hbar\omega}}$ for several important quantum Fisher information are summarized in Table.~\ref{table}.

\begin{table}%[H] %add [H] placement to break table across pages
\caption{List of the coefficients appearing in the generalized fluctuation-dissipation theorem~\eqref{gFDT} for various quantum Fisher information. \label{table}}
\begin{ruledtabular}
\begin{tabular}{lcc}
quantum Fisher information&$f(x)$&$\frac{f(\ex^{-\alpha})}{1-\ex^{-\alpha}}$\\\hline\hline
symmetric logarithmic & $\frac{x+1}{2}$ & $\frac{1}{2}\coth\frac{\alpha}{2}=\bar{n}+1/2$ \\
derivative &  &  \\\hline
Bogoliubov-Kubo-Mori & $\frac{x-1}{\log x}$ & $\frac{1}{\alpha} = \frac{1}{\log (\bar{n}+1)-\log\bar{n}}$\\\hline
right logarithmic derivative& $x$ & $\frac{1}{\ex^\alpha-1} =\bar{n}$  \\\hline
left logarithmic derivative& $1$ & $\frac{1}{1-\ex^{-\alpha}}=\bar{n}+1$  \\\hline
real part of right logarithmic& $\frac{2x}{x+1}$ & $\frac{1}{\sinh\alpha}=\frac{2\bar{n}(\bar{n}+1)}{2\bar{n}+1}$  \\
 derivative&  &  \\\hline
skew information&$\frac{(\sqrt x +1)^2}{4}$&$\frac{1}{4}\coth\frac{\alpha}{4}=\frac{2\bar{n}+1+\sqrt{\bar{n}(\bar{n}+1)}}{4}$\\
\end{tabular}
\end{ruledtabular}
\end{table}

%------------------------------------------------------------------------------------%
%------------------------------------------------------------------------------------%
The crucial step in deriving Eq.~\eqref{gFDT} is to write down the complicated action of the superoperator $\bm K^f_{\h\rho}$ with c-numbers by considering the matrix components in the basis of $\ket i$'s, as
\e{
\braket{j|\bm K^{f}_{\h\rho}(\h J_\nu)|i}=p_i f\left( \frac{p_j}{p_i} \right)\braket{j|\h J_\nu|i}.
}
Then the Fourier transform of the generalized covariance can be calculated as 
%\begin{widetext}
\e{
C_{\mu\nu,\omega}^f&=2\pi\hbar f(\ex^{-\beta\hbar\omega})\\
&\quad \times \sum_{i,j} \delta(E_i-E_j-\hbar\omega)p_i \braket{i|\h J_\mu|j}\braket{j|\h J_\nu|i},\label{FGC}
}
%\end{widetext}
where we have used the fact that $E_j-E_i=\hbar\omega$ and hence $p_j/p_i=\ex^{-\beta\hbar\omega}$ due to the existence of the $\delta$ function.
A similar calculation leads to the expression of the Fourier transform of the response function $\Phi_{\mu\nu,\omega}$ from Eq.~\eqref{Kubo} as 
\e{
&\Phi_{\mu\nu,\omega}\nonumber\\
=& 2\pi\hbar \frac{1-\ex^{-\beta\hbar\omega}}{\hbar\omega}\sum_{i,j} \delta(E_i-E_j-\hbar\omega)p_i \braket{i|\h J_\mu|j}\braket{j|\h J_\nu|i} . \label{FLF}
}
Comparing Eqs.~\eqref{FGC} and~\eqref{FLF}, we obtain Eq.~\eqref{gFDT}.
See the Supplemental Material for the more detailed derivation of the generalized fluctuation-dissipation theorem~\eqref{gFDT}.

%------------------------------------------------------------------------------------%
%------------------------------------------------------------------------------------%
By using the generalized fluctuation-dissipation theorem~\eqref{gFDT}, we can reconstruct the generalized covariance from the admittance
\e{
\chi_{\mu\nu}(\omega):=\int^\infty_0 {\rm d}t\ex^{i\omega t}\Phi_{\mu\nu}(t),
}
which is experimentally measurable by a following procedure.
When we add a harmonically oscillating external force $X_\nu(t)={\rm Re}[X_\nu \ex^{i\omega t}]$, the expectation value of the current also oscillates as $\braket{\h J_\mu(t)}={\rm Re}[\chi_{\mu\nu,\omega}X_\nu \ex^{i\omega t}]$.
Therefore, it is sufficient to measure the amplitude and the phase of the oscillation of the current to determine the admittance.
Then, the inverse Fourier transform of Eq.~\eqref{gFDT} with $\omega=0$ gives
\e{
\ginprod{\h J_\mu,\h J_\nu}{\h\rho}{f}=\int^\infty_{-\infty}\frac{{\rm d}\omega}{2\pi}\frac{\hbar\omega f(\ex^{-\beta\hbar\omega})}{1-\ex^{-\beta\hbar\omega}} \left(\chi_{\mu\nu}(\omega)+\chi_{\nu\mu}(\omega)^*\right).\label{gcm}
}
Therefore, all the types of generalized covariances can be determined by measuring the response to the harmonically oscillating external perturbation for all frequencies $0<\omega<\infty$.
Furthermore, if the generalized covariance is symmetric, i.e., $f(x)$ is standard, then Eq.~\eqref{gcm} can be simplified, giving
\e{
\ginprod{\h J_\mu,\h J_\nu}{\h\rho}{f}\overset{f=\tilde{f}}{=}\frac{2}{\pi} \int^\infty_0{\rm d}\omega \frac{\hbar\omega  f(\ex^{-\beta\hbar\omega})}{1-\ex^{-\beta\hbar\omega}}{\rm Re}[\chi_{\mu\nu}^{\rm s}(\omega)  ],\label{gcmsimp}
}
where $\chi_{\mu\nu}^{\rm s}(\omega):=(\chi_{\mu\nu}(\omega)+\chi_{\nu\mu}(\omega))/2$ is a symmetric part of the admittance matrix.

%------------------------------------------------------------------------------------%
%------------------------------------------------------------------------------------%
Similar calculations lead to the generalized fluctuation-dissipation theorem for the displacement operator $\h A_\mu$.
In fact, if we define the Fourier transform of the covariance of two temporal displacement operators as
\e{
\tilde{C}_{\mu\nu,\omega}^f&:=\int_{-\infty}^\infty{\rm d}(t-t')\ex^{i(t-t')\omega}\ginprod{\Delta\h A_\mu(t), \Delta\h A_\nu(t')}{\h\rho}{f},
}
we can show that it is also proportional to the linear response function as
\e{
\tilde{C}_{\mu\nu,\omega}^f&=-i\hbar\frac{f(\ex^{-\beta\hbar\omega})}{1-\ex^{-\beta\hbar\omega}}\tilde{\Phi}_{\mu\nu,\omega}.
}
Therefore, we obtain the expression of the generalized covariance of the two displacement operators in terms of the dynamical susceptibility $\tilde{\chi}_{\mu\nu}(\omega):=\int^\infty_0 {\rm d}t\ex^{i\omega t}\tilde{\Phi}_{\mu\nu}(t)$,
as
\e{
\ginprod{\Delta\h A_\mu,\Delta\h A_\nu}{\h\rho}{f}&= \frac{i}{\hbar}\int^\infty_{-\infty} \frac{{\rm d}\omega}{2\pi} \frac{f(\ex^{-\beta\hbar\omega})}{1-\ex^{-\beta\hbar\omega}} \nonumber\\
&   \qquad\quad\qquad \times(\tilde{\chi}_{\mu\nu}(\omega) - \tilde{\chi}_{\nu\mu}(\omega)^* ) .
}
If $f(x)=\tilde{f}(x)$, we can simplify the formula as
\e{
\ginprod{\Delta\h A_\mu,\Delta\h A_\nu}{\h\rho}{f} \overset{f=\tilde{f}}{=} \frac{2\hbar}{\pi}\int^\infty_0 {\rm d}\omega \frac{f(\ex^{-\beta\hbar\omega})}{1-\ex^{-\beta\hbar\omega}} {\rm Im}[\tilde{\chi}_{\mu\nu}^s(\omega)]. \label{gcmsimp2}
}

Thus, once we  measure the admittance or the dynamical susceptibility for all frequencies, we can determine any type of generalized covariance through the simple post-processing~\eqref{gcmsimp} or~\eqref{gcmsimp2}.

%------------------------------------------------------------------------------------%
%------------------------------------------------------------------------------------%
\section{determining the quantum Fisher information}\label{determine}
The quantum Fisher information can also be determined via Eqs.~\eqref{gcm} or~\eqref{gcmsimp} because it is nothing but the generalized covariance of the logarithmic derivative~\eqref{QFIGC}.

Here, we explicitly calculate the external field that we need to apply for a specific model.
Consider a one-parameter model which is given by
\e{
\h\rho_\theta:=\ex^{-i\theta\h B}\h\rho\ex^{i\theta\h B}.\label{model}
}
The parameter $\theta$ to be estimated is the degree of shift or rotation, and $\h B$ is the generator.
Then, the quantum Fisher information of this model at $\theta=0$ is equal to the generalized covariance~\eqref{gcm} if the perturbation $\h A$ is chosen so that the corresponding current operator $\h J$ coincides with the logarithmic derivative $\h L:=(\bm K^f_{\h\rho})^{-1}(i[\h\rho, \h B])$.
In information geometry~\cite{Amari2007}, the logarithmic derivative $\h L_\mu$ is interpreted as the e-representation of a tangent vector $\partial/\partial \theta_\mu$ on the quantum state space.
Solving the equation $\h J=\h L$, we obtain the matrix element of the external  field $\h A$ as
\e{
 E_{ij}\frac{f(\ex^{-\beta E_{ij}})}{1-\ex^{-\beta E_{ij}}} \braket{i|\h A|j}= \braket{i|\h B|j},
}
where $ E_{ij}:=E_i-E_j$ is the energy difference.
It is worth noting that the ratio of the matrix elements is equal to the coefficient appeared in Eq.~\eqref{gFDT}.

When we are able to measure the dynamical susceptibility, we can take the perturbation $\h A$ to be the very generator $\h B$.
In fact, the quantum Fisher information of the unitary model~\eqref{model} corresponding to the operator monotone function $f(x)$ can be expressed as the generalized covariance corresponding to $(x-1)^2/f(x)$:
\e{
J^Q_{\theta=0}&=\sum_{i,j}p_j\frac{(1-p_i/p_j)^2}{f(p_i/p_j)}\big|\braket{i|\h A|j}\big|^2\\
&=\ginprod{\Delta\h A,\Delta\h A}{\h\rho}{(x-1)^2/f(x)}. \label{gc-fisher}
}
Therefore, we obtain
\e{
J^Q_{\theta=0}&=\frac{2\hbar}{\pi}\int^\infty_0 {\rm d}\omega \frac{1-\ex^{-\beta\hbar\omega}}{f(\ex^{-\beta\hbar\omega})} {\rm Im}[\tilde{\chi}(\omega)],\label{Hauke}
}
where $\tilde{\chi}(\omega)$ is the dynamical susceptibility of $\h A=\h B$ when perturbation $\h B$ is applied.
If we set $f(x)=(x+1)/2$, Eq.~\eqref{Hauke} reduces to the previous study~\cite{Hauke2016} for the SLD Fisher information.

%------------------------------------------------------------------------------------%
%------------------------------------------------------------------------------------%
\section{skew information and uncertainty relation}\label{SIUR}
Skew information was first introduced by Wigner and Yanase~\cite{Wigner1963} to express the information contained by the quantum state when a conserved quantity exists, and is defined by
\e{
I_{1/2}(\h\rho,\h A):=-\frac{1}{2}\tr{\left([\h\rho^{1/2},\h A]\right)^2}.
}
Dyson made a one-parameter extension of the skew information, which is called Wigner-Yanase-Dyson (WYD) skew information, as
\e{
I_\alpha(\h\rho,\h A)&:=-\frac{1}{2}\tr{[\h\rho^\alpha,\h A][\h\rho^{1-\alpha},\h A]}.
}
In fact, the WYD skew information is quantitatively related to the quantum Fisher information of the unitary model~\eqref{model}~\cite{GIbilisco2003,Hansen2008} as
\e{
I_\alpha(\h\rho,\h A)=\frac{\alpha(1-\alpha)}{2} J^Q_{\theta=0},
}
where the operator monotone function of the quantum Fisher information is chosen to be  
\e{
f_\alpha(x)&=\alpha(1-\alpha)\frac{(x-1)^2}{(x^\alpha-1)(x^{1-\alpha}-1)}. \label{WYDfunction}
}
Based on this observation, Hansen further generalized it to the metric adjusted skew information~\cite{Hansen2008}
\e{
I_f(\h\rho,\h A)&:=\frac{f(0)}{2} J^Q_{\theta=0},
}
where $f(x)$ is an arbitrary operator monotone function satisfying $f(0)\ne 0$.

Since the metric adjusted skew information satisfies some desirable properties such as the convexity with respect to the state~\cite{Hansen2008}, it can be regarded as a quantum part of the fluctuation of the observable $\h A$ in the mixed state $\h\rho$.
Therefore, it has been used to formulate uncertainty relations~\cite{Luo2005,Gibilisco2007a,Yanagi2010,Furuichi2010,Yanagi2011} for a quantum state, which is tighter than the conventional uncertainty relation~\cite{Robertson1934}.
However, since the skew information includes the term such as $\tr{\h\rho^\alpha\h A\h\rho^{1-\alpha}\h A}$, it cannot be measured by usual quantum measurements.
Our method developed in Sec.~\ref{determine} provides a way to experimentally determine the skew information and hence all the quantities used in various formulations of skew information-based uncertainty relations~\cite{Luo2005,Gibilisco2007a,Yanagi2010,Furuichi2010,Yanagi2011}.
The metric adjusted skew information can be determined by the formula
\e{
 I_f(\h\rho,\h A)&=\frac{f(0)\hbar}{\pi}\int^\infty_0 {\rm d}\omega \frac{1-\ex^{-\beta\hbar\omega}}{f(\ex^{-\beta\hbar\omega})} {\rm Im}[\tilde{\chi}(\omega)],\label{formulaskew}
}
where $\tilde{\chi}(\omega)$ is the dynamical susceptibility of $\h A$ when the external perturbation $\h A$ is applied.

As a concrete example, we apply our result to a harmonic oscillator system in thermal equilibrium, and demonstrate that the uncertainty relation shown in Ref.~\cite{Yanagi2010} can be verified experimentally.
Let us define the quantum fluctuation of the observable $\h A$ by
\e{
U_\alpha(\h\rho,\h A):=\sqrt{\braket{(\Delta\h A)^2}_{\h\rho}^2 - \left(\braket{(\Delta\h A)^2}_{\h\rho}- I_\alpha(\h\rho,\h A) \right)^2 }.
}
Yanagi~\cite{Yanagi2010} showed that the uncertainty relation
\e{
U_\alpha(\h\rho,\h A)U_\alpha(\h\rho,\h B)\ge \alpha(1-\alpha)\big|\tr{\h\rho[\h A,\h B]}\big|^2   \label{WYDuncertainty}
}
holds for any $\alpha\in[0,1]$.
We consider a harmonic oscillator in thermal equilibrium:
\e{
\h H=\frac{1}{2m}\h p^2 + \frac{1}{2}m\omega^2\h x^2 .
}
We apply an external perturbation corresponding to the position and the momentum operators
\e{
\h A_\mu:=\begin{cases}
    \h x & (\mu=1); \\
    \h p & (\mu=2).
  \end{cases}
}
Then, the diagonal components of the dynamical susceptibility become
\e{
&m\omega\tilde{\chi}_{11}(\omega')=\frac{1}{m\omega}\tilde{\chi}_{22}(\omega')\nonumber \\
=&\frac{1}{2}\left( {\mathcal P}\frac{1}{\omega'+\omega} -{\mathcal P} \frac{1}{\omega'-\omega} -i\pi\delta(\omega'+\omega) + i\pi\delta(\omega'-\omega)    \right),
}
where $\mathcal P$ denotes the principal value.
Therefore, from Eqs.~\eqref{WYDfunction} and~\eqref{formulaskew}, the WYD skew information is calculated to be
\e{
I_\alpha(\h\rho,\h x)&=\frac{\hbar}{2m\omega}\cdot\frac{(1-\ex^{-\alpha\beta\hbar\omega})(1- \ex^{-(1-\alpha)\beta\hbar\omega})}{1-\ex^{-\beta\hbar\omega}},\\
I_\alpha(\h\rho,\h p)&=\frac{\hbar m\omega}{2}\cdot\frac{(1-\ex^{-\alpha\beta\hbar\omega})(1- \ex^{-(1-\alpha)\beta\hbar\omega})}{1-\ex^{-\beta\hbar\omega}},
}
and the inequality~\eqref{WYDuncertainty} reduces to
\e{
\frac{(1-\ex^{-2\alpha\beta\hbar\omega})(1-\ex^{-2(1-\alpha)\beta\hbar\omega})}{(1-\ex^{-\beta\hbar\omega})^2}\ge 4\alpha(1-\alpha).
\label{HOuncertainty}}

It is worth noting that the equality in Eq.~\eqref{HOuncertainty} holds when $\alpha=1/2$ even though the state is the mixed state and includes non-minimum uncertainty states.
Here, we examine the simple harmonic oscillator system so that the analytical calculation of the dynamical susceptibility can be performed.
Our method enables us to experimentally determine the information of the complex system for which the analytical treatment is difficult.

%------------------------------------------------------------------------------------%
%------------------------------------------------------------------------------------%
\section{discussion}\label{discussion}
We discuss the applicability and the efficiency of the proposed method.
First, our method is applicable in two situations: the Hamiltonian is given and we want to know the quantum Fisher information of the thermal equilibrium state under that Hamiltonian; the density operator is given and we want to know the quantum Fisher information of that state.
For the latter case, an effective Hamiltonian $\h H=-\frac{1}{\beta}\log\h\rho$ tells us what kind of Hamiltonian we need to engineer.
Such a situation seems realistic when the system size is relatively small.
Second, the estimation via the integral~\eqref{gcm} is efficient for the following reason.
The integrand in Eq.~\eqref{gcm} consists of the delta functions that contribute only if the frequency matches the energy difference of two eigenstates, and hence the integral can be rewritten in terms of a discrete sum.
When the system is small (e.g., a few qubits), the number of measurement required to estimate the sum is also small.
As the size of the system becomes large, the number of the terms in the sum becomes exponentially large.
For such a large system, however, the integrand in Eq. (21) can be approximated as a continuous function. Therefore, the error of the estimate can be controlled by the space of sampling frequencies, and does not depend on the system size.

%------------------------------------------------------------------------------------%
%------------------------------------------------------------------------------------%
\section{conclusion}\label{conclusion}
In summary, we have derived the generalized fluctuation-dissipation theorem~\eqref{gFDT} which relates the response function to generalized covariances.
On the basis of this theorem, we have developed a method to determine  the generalized covariance from the admittance, which can be experimentally determined by applying an ac external field.
We have also identified the direction of the external field needed to determine the quantum Fisher information.
Our results thus point to an interesting connection between statistical mechanics and the Fisher information in the quantum regime through the generalized covariance.

\if0
\section{NOTE ADDED}
After completion of our work, we became aware of the paper~\cite{Hauke2016}, which derives a relation between the SLD Fisher information and the susceptibility similar to our result~\eqref{gcmsimp}.
%We note that their result is easily generalized to any kind of the quantum Fisher metric, but different from ours since the external field to be applied is different.
\fi

\begin{acknowledgments}
This work was supported by KAKENHI Grant No. 26287088 from the Japan Society for the Promotion of Science, Grant-in-Aid for Scientific Research on Innovation Areas ``Topological Materials Science'' (KAKENHI Grant No. 15H05855), the Photon Frontier Network Program from MEXT of Japan, and the Mitsubishi Foundation.
T. S. was supported by the Japan Society for the Promotion of Science through Program for Leading Graduate Schools (ALPS).

\end{acknowledgments}

\appendix
\section{derivation of Eq.~\eqref{gFDT}}
We assume that the state is described by the canonical ensemble, so that
\e{
\h H&=\sum_i E_i \ket i\bra i,\\
\h\rho&=\sum_i p_i\ket i\bra i,
}
where $p_i=\ex^{-\beta(E_i-F)}$, and $F=-\frac{1}{\beta}\log\tr{\ex^{-\beta\h H}}$ is the free energy.
First, we calculate the action of the superoperator $\bm K^f_{\h\rho}=f(\bm L_{\h\rho}\bm R_{\h\rho}^{-1})\bm R_{\h\rho}$ that appears in the definition of the generalized covariance.
If $f(x)=x^k$, we obtain
\e{
\braket{j|\bm K^{f(x)=x^k}_{\h\rho}\h J_\nu|i}&=\braket{j|\h\rho^k\h J_\nu\h\rho^{1-k}|i}\nonumber\\
&=p_i\left( \frac{p_j}{p_i} \right)^k \braket{j|\h J_\nu|i}.
}
Therefore, for a polynomial $f(x)=\sum_{k=1}^n c_kx^k$, we obtain
\e{
\braket{j|\bm K^{f}_{\h\rho}\h J_\nu|i}&=\sum_{k=1}^nc_kp_i\left( \frac{p_j}{p_i} \right)^n \braket{j|\h J_\nu|i}\nonumber\\
&=p_i f\left( \frac{p_j}{p_i} \right)\braket{j|\h J_\nu|i}. \label{superoperatoraction}
}
Let $m$ and $M$ be the minimum and maximum of $p_i/p_j$'s, respectively.
Since the operator monotone function $f(x)$ is continuous~\cite{Bhatia1997}, it can be uniformly approximated by polynomials on the closed interval $[m,M]$ from the Weierstrass approximation theorem.
Therefore, the relation~\eqref{superoperatoraction} holds for an arbitrary operator monotone function $f(x)$.
Then, the generalized covariance can be calculated as follows:
%\begin{widetext}
\e{
&C^f_{\mu\nu,\omega} \nonumber\\
=&\int_{-\infty}^\infty {\rm d}t \ex^{i\omega t} \ginprod{\h J_\mu(t),\h J_\nu(0)}{\h\rho}{f}\nonumber\\
=&\int_{-\infty}^\infty {\rm d}t \ex^{i\omega t} \sum_{i,j} \braket{i|\ex^{i\h Ht/\hbar}\h J_\mu\ex^{-i\h Ht/\hbar}|j}\braket{j|\bm K^{f}_{\h\rho}\h J_\nu|i}\nonumber\\
=&\sum_{i,j} \int_{-\infty}^\infty {\rm d}t\ex^{i(\hbar\omega+E_i-E_j)t/\hbar}p_if\left( \frac{p_j}{p_i} \right) \braket{i|\h J_\mu|j}\braket{j|\h J_nu|i}\nonumber\\
=&\sum_{i,j}2\pi\hbar\delta(\hbar\omega+E_i-E_j)p_if\left( \frac{p_j}{p_i} \right) \braket{i|\h J_\mu|j}\braket{j|\h J_nu|i}\nonumber\\
=&2\pi\hbar f(\ex^{-\beta\hbar\omega}) \sum_{i,j} \delta(\hbar\omega+E_i-E_j)p_i \braket{i|\h J_\mu|j}\braket{j|\h J_\nu|i}.  \label{covariancefrequency}
}
%\end{widetext}
In deriving the last equality, we use the fact that $E_j-E_i=\hbar\omega$ and hence $p_j/p_i=\ex^{-\beta\hbar\omega}$ due to the $\delta$ function.

Next, we calculate the response function as
%\begin{widetext}
\e{
&\Phi_{\mu\nu}(t)  \nonumber\\
=&\beta\int^1_0{\rm d}\lambda\tr{\h\rho^\lambda \h J_\mu(t) \h\rho^{1-\lambda} \h J_\nu(0)}\nonumber\\
=&\beta \int^1_0 {\rm d}\lambda\sum_{i,j} \braket{i|\h\rho^\lambda \ex^{i\h Ht/\hbar}\h J_\mu\ex^{-i\h Ht/\hbar} |j}\braket{j|\h\rho^{1-\lambda} \h J_\nu|i}\nonumber\\
=&\beta\int^1_0{\rm d}\lambda\sum_{i,j} \ex^{i(E_i-E_j)t/\hbar} p_j \left( \frac{p_i}{p_j} \right)^\lambda \braket{i|\h J_\mu|j}\braket{j|\h J_\nu|i}\nonumber\\
=&\beta \sum_{i,j} \ex^{i(E_i-E_j)t/\hbar}  \frac{p_i-p_j}{\log p_i-\log p_j}  \braket{i|\h J_\mu|j}\braket{j|\h J_\nu|i}.
}
%\end{widetext}
By Fourier transforming it, we obtain
%\begin{widetext}
\e{
&\Phi_{\mu\nu,\omega}\nonumber\\
=&\int_{-\infty}^\infty {\rm d}t \ex^{i\omega t}\Phi_{\mu\nu}(t)\nonumber\\
=&\beta\sum_{i,j} 2\pi\hbar\delta(E_i-E_j+\hbar\omega)\nonumber\\
&\quad \quad \times p_i\frac{1-p_j/p_i}{-\log(p_j/p_i)} \braket{i|\h J_\mu|j}\braket{j|\h J_\nu|i}\nonumber\\
=&\beta 2\pi\hbar \frac{1-\ex^{-\beta\hbar\omega}}{\beta\hbar\omega}\sum_{i,j} \delta(E_i-E_j+\hbar\omega)p_i \braket{i|\h J_\mu|j}\braket{j|\h J_\nu|i}\nonumber\\
=&2\pi\hbar \frac{1-\ex^{-\beta\hbar\omega}}{\hbar\omega}\sum_{i,j} \delta(E_i-E_j+\hbar\omega)p_i \braket{i|\h J_\mu|j}\braket{j|\h J_\nu|i},\label{responsefrequency}
}
%\end{widetext}
where we have use $p_j/p_i=\ex^{-\beta\hbar\omega}$ to obtain the second last equality.
Comparing Eq.~\eqref{covariancefrequency} and Eq.~\eqref{responsefrequency}, we obtain the generalized fluctuation-dissipation theorem, namely,
\e{
C^f_{\mu\nu,\omega}=\frac{\hbar\omega f(\ex^{-\beta\hbar\omega})}{1-\ex^{-\beta\hbar\omega}}{\Phi_{\mu\nu,\omega}}.
}
\bibliography{/users/shitaratomohiro/Documents/library.bib}

\end{document}